\documentclass[11pt,twocolumn]{aastex63}

\shorttitle{Density Fluctuations}
\shortauthors{Krupar et al.}
\received{September 13, 2019}
\accepted{December 22, 2019}
\submitjournal{ApJS}

\begin{document}

\title{Density Fluctuations in the Solar Wind Based on Type III Radio Bursts Observed by Parker Solar Probe}

\correspondingauthor{Vratislav Krupar}
\email{vratislav.krupar@nasa.gov,vk@ufa.cas.cz}

\author[0000-0001-6185-3945]{Vratislav Krupar}
\affil{Universities Space Research Association, Columbia, MD 21046, USA}
\affil{Heliospheric Physics Laboratory, Heliophysics Division, NASA Goddard Space Flight Center, Greenbelt, MD 20771, USA}
\affil{Department of Space Physics, Institute of Atmospheric Physics of the Czech Academy of Sciences, Prague 14131, Czech Republic}

\author[0000-0003-3255-9071]{Adam Szabo}
\affil{Heliospheric Physics Laboratory, Heliophysics Division, NASA Goddard Space Flight Center, Greenbelt, MD 20771, USA}

\author[0000-0001-6172-5062]{Milan Maksimovic}
\affil{LESIA, Observatoire de Paris, Universit\'{e} PSL, CNRS, Sorbonne Universit\'{e}, Universit\'{e} de Paris, 92195 Meudon, France}

\author[0000-0002-1122-6422]{Oksana Kruparova}
\affil{Department of Space Physics, Institute of Atmospheric Physics of the Czech Academy of Sciences, Prague 14131, Czech Republic}

\author[0000-0002-8078-0902]{Eduard P. Kontar}
\affil{School of Physics and Astronomy, University of Glasgow, Glasgow G12 8QQ, UK}

\author[0000-0003-1162-5498]{Laura A. Balmaceda}
\affil{Heliospheric Physics Laboratory, Heliophysics Division, NASA Goddard Space Flight Center, Greenbelt, MD 20771, USA}
\affil{George Mason University, Arlington, VA 22030, USA}

\author[0000-0003-4217-7333]{Xavier Bonnin}
\affil{LESIA, Observatoire de Paris, Universit\'{e} PSL, CNRS, Sorbonne Universit\'{e}, Universit\'{e} de Paris, 92195 Meudon, France}


\author[0000-0002-1989-3596]{Stuart D. Bale}
\affil{Physics Department, University of California, Berkeley, CA 94720-7300, USA}
\affil{Space Sciences Laboratory, University of California, Berkeley, CA 94720-7450, USA}
\affil{The Blackett Laboratory, Imperial College London, London, SW7 2AZ, UK}
\affil{School of Physics and Astronomy, Queen Mary University of London, London E1 4NS, UK}

\author[0000-0002-1573-7457]{Marc Pulupa}
\affil{Space Sciences Laboratory, University of California, Berkeley, CA 94720-7450, USA}

\author[0000-0003-1191-1558]{David M. Malaspina}
\affil{Laboratory for Atmospheric and Space Physics, University of Colorado, Boulder, CO 80303, USA}

\author[0000-0002-0675-7907]{John W. Bonnell}
\affil{Space Sciences Laboratory, University of California, Berkeley, CA 94720-7450, USA}

\author[0000-0002-6938-0166]{Peter R. Harvey}
\affil{Space Sciences Laboratory, University of California, Berkeley, CA 94720-7450, USA}

\author[0000-0003-0420-3633]{Keith Goetz}
\affiliation{School of Physics and Astronomy, University of Minnesota, Minneapolis, MN 55455, USA}

\author[0000-0002-4401-0943]{Thierry {Dudok de Wit}}
\affil{LPC2E, CNRS and University of Orl\'eans, Orl\'eans, France}

\author[0000-0003-3112-4201]{Robert J. MacDowall}
\affil{Solar System Exploration Division, NASA Goddard Space Flight Center, Greenbelt, MD 20771, USA}

\author[0000-0002-7077-930X]{Justin C. Kasper}
\affiliation{Climate and Space Sciences and Engineering, University of Michigan, Ann Arbor, MI 48109, USA}
\affiliation{Smithsonian Astrophysical Observatory, Cambridge, MA 02138, USA}

\author[0000-0002-3520-4041]{Anthony W. Case}
\affiliation{Smithsonian Astrophysical Observatory, Cambridge, MA 02138, USA}

\author[0000-0001-6095-2490]{Kelly E. Korreck}
\affiliation{Smithsonian Astrophysical Observatory, Cambridge, MA 02138, USA}

\author[0000-0001-5030-6030]{Davin E. Larson}
\affil{Space Sciences Laboratory, University of California, Berkeley, CA 94720-7450, USA}

\author[0000-0002-0396-0547]{Roberto Livi}
\affil{Space Sciences Laboratory, University of California, Berkeley, CA 94720-7450, USA}

\author{Michael L. Stevens}
\affiliation{Smithsonian Astrophysical Observatory, Cambridge, MA 02138, USA}

\author[0000-0002-7287-5098]{Phyllis L. Whittlesey}
\affil{Space Sciences Laboratory, University of California, Berkeley, CA 94720-7450, USA}

\author{Alexander M. Hegedus}
\affiliation{Climate and Space Sciences and Engineering, University of Michigan, Ann Arbor, MI 48109, USA}

\begin{abstract}

Radio waves are strongly scattered in the solar wind, so that their apparent sources seem to be considerably larger and shifted than the actual ones. Since the scattering depends on the spectrum of density turbulence, better understanding of the radio wave propagation provides indirect information on the relative density fluctuations $\epsilon=\langle\delta n\rangle/\langle n\rangle$ at the effective turbulence scale length. Here, we have analyzed 30 type III bursts detected by Parker Solar Probe (PSP).
\replaced{to retrieve decay times $\tau_{\rm{d}}$ as a function of frequency $f$.}{For the first time, we have retrieved type III burst decay times $\tau_{\rm{d}}$ between $1$~MHz and $10$~MHz thanks to an unparalleled temporal resolution of PSP.}
We observed a significant deviation in a power-law slope for frequencies above 1 MHz when compared to previous \replaced{observations}{measurements below $1$~MHz} by the twin-spacecraft Solar TErrestrial RElations Observatory (\textit{STEREO}) mission. \replaced{Next, we performed Monte Carlo simulations to study the role of scattering on time-frequency profiles of radio emissions.}{We note that altitudes of radio bursts generated at 1 MHz roughly coincide with an expected location of the Alfv\'{e}n point, where the solar wind becomes super-Alfv\'{e}nic.}
By comparing PSP observations and Monte Carlo simulations we \replaced{predicted}{predict} relative density fluctuations $\epsilon$ at the effective turbulence scale length at radial distances between 2.5~$R_\Sun$ and 14~$R_\Sun$ to range from $0.22$ and $0.09$. Finally, we calculated relative density fluctuations $\epsilon$ measured \textit{in situ} by PSP at a radial distance from the Sun of $35.7$~$R_\Sun$ during the perihelion \#1, and the perihelion \#2 to be $0.07$ and $0.06$, respectively. It is in a very good agreement with previous \textit{STEREO} predictions  ($\epsilon=0.06-0.07$) obtained by remote measurements of radio sources generated at this radial distance.

\end{abstract}


\section{Introduction}\label{sec:introduction}
Type III bursts belong among the strongest radio signals routinely observed by both\replaced{ spacecraft and ground based}{, space-borne and ground-based} observatories \citep{1998ARA&A..36..131B,2017JSWSC...7A..37M}.
They are generated by electron beams accelerated at reconnection sites of solar flares traveling outward along open magnetic filed lines through the corona and the interplanetary medium \citep{1950AuSRA...3..541W}.
Along their path electrons beams interact  with the background plasma  producing radio emissions at the electron plasma frequency $f_{\rm{pe}}$ (the fundamental component), and/or at its first harmonic $2f_{\rm{pe}}$ (the harmonic component) via the plasma emission mechanism \citep{1958SvA.....2..653G,1995ApJ...453..959C,1998ApJ...503..435E}.
\added{Generally, the two components can be distinguished when observed simultaneously, which is rather typical at decametric or shorter wavelengths \citep{1974SoPh...39..451S}.
The fundamental component is usually more intense with 2-3 times higher circular polarization \citep{1980A&A....88..203D}.
However, it is almost impossible to distinguish the two components in time and frequency, or by polarization for type III bursts at longer wavelengths, which are generated in the interplanetary medium \citep{1998JGR...10329651R,2005JGRA..11012S07G,2015A&A...580A.137K}.
For rare cases when electron beams are detected \textit{in situ} at the spacecraft, the initial radiation is almost always the fundamental component, while in the late phases it may be one or another \citep{1998JGR...10317223D}.
Currently, there is no observational evidence to choose between the fundamental and harmonic component for interplanetary type III bursts.}

Type III bursts can be simultaneously detected over a broad range of \replaced{angles}{longitudes}, even if their sources are located behind the Sun \citep{2008A&A...489..419B}. Their apparent radio sources lie at considerably larger radial distances than predicted by electron density models \citep{2009SoPh..259..255R,2012SoPh..279..153M}. Furthermore, apparent type III burst source sizes are so extended that may spread over the entire inner heliosphere \citep{2014SoPh..289.4633K}. These obscure properties are attributed to scattering of radio waves by electron density inhomogeneities as they propagate from the source region to the observer \citep{1984A&A...140...39S,1985A&A...150..205S,1994ApJ...426..774B,1995ApJ...439..494B,2018NatCo...9..146K}.
The role of refraction and scattering of interplanetary radio emissions can be studied using a geometric optics method and Monte Carlo simulations \citep{1968AJ.....73..972H,1980SSRv...26....3M,2007ApJ...671..894T,2008ApJ...676.1338T,2019ApJ...884..122K}.

Recently, \citet{2018ApJ...857...82K} compared decay times of type III bursts between $125$~kHz and $1$~MHz observed by the \replaced{\textit{STEREO}}{Solar TErrestrial RElations Observatory (\textit{STEREO})} spacecraft with results of Monte Carlo simulations. They suggest that the characteristic exponential decay profile of type III bursts could be solely explained by the scattering of the fundamental component between the source and the observer.
\citet{2018ApJ...857...82K} estimated relative electron density fluctuations $\epsilon=\langle\delta n\rangle/\langle n\rangle$ to be $0.06$--$0.07$ at radial distances from the Sun between $8$ and $45$~solar radii (1~$R_\Sun$~=~695,500~km), where $\langle n \rangle$ represents an average electron density and $\langle \delta n \rangle$ is an average amplitude of variations of an electron density $n$ from its average value $\langle n \rangle$.

Here, we primarily examine radio measurements obtained by the Parker Solar Probe (PSP) mission with a
perihelion down to $9.5$~$R_\Sun$ and aphelion near $\sim1$~astronomical unit \citep[$1$~au = 149,598,000~km; ][]{2016SSRv..204....7F}.
The PSP/FIELDS instrument provides with comprehensive measurements of coronal plasma and magnetic field, plasma waves and turbulence, and radio signatures of solar transients \citep{2016SSRv..204...49B,2017JGRA..122.2836P}.
We use data recorded by the Radio Frequency Spectrometer (RFS), which is a two-channel digital receiver and spectrometer in the FIELDS suite.
Specifically, we analyze time-frequency profiles of type III bursts between $0.5$~MHz and $10$~MHz recorded by the RFS/Low Frequency Receiver (LFR; \added{64 logarithmically spaced frequency channels between}
$10.5$~kHz -- $1.7$~MHz \added{with a temporal resolution of $7$~s}) and the RFS/High Frequency Receiver (HFR;
\added{64 logarithmically spaced frequency channels between} $1.3$~MHz -- $19.2$~MHz
\added{with a temporal resolution of $7$~s}). Frequencies above $10$~MHz have been excluded from this study due to insufficient time resolution of RFS/HFR\deleted{ which is $7$~seconds}.
On the other hand the frequencies below $0.5$~MHz are strongly affected by the quasi-thermal noise \replaced{\citep{1989JGR....94.2405M}}{\citep[QTN; ][]{1989JGR....94.2405M} due to considerably larger solar wind density near the Sun when compared to $1$~au}.
For a case study we also use radio data recorded by the \textit{Wind}/WAVES and \replaced{Solar TErrestrial RElations Observatory (\textit{STEREO})}{\textit{STEREO}}/WAVES instruments \added{with a temporal resolution of $60$~s and $35$~s, respectively} \citep{1995SSRv...71..231B,2008SSRv..136..487B}.
Finally, we investigate solar wind density and bulk velocity retrieved by The \added{PSP/}Solar Wind Electrons Alphas and Protons (SWEAP) instrument \citep{2016SSRv..204..131K}.

In this paper, we present a statistical survey of type III burst decay times that can be used to estimate relative electron density fluctuations $\epsilon$ in the solar wind. In Section~\ref{sec:obs} we present our analysis of RFS measurements (Section~\ref{sec:mes}), its comparison to results of Monte Carlo simulations (Section~\ref{sec:sim}), and relative density fluctuations $\epsilon$ measured in situ by PSP (Section~\ref{sec:fluc}). \replaced{The concluding remarks are listed}{Finally, we discuss and summarize our findings} in Section~\ref{sec:con}.

\section{Observation and Analysis}
\label{sec:obs}

\subsection{Type III Bursts Measurements}
\label{sec:mes}

\begin{figure*}[htb]
\centering
\includegraphics[angle=0,width=0.8\linewidth]{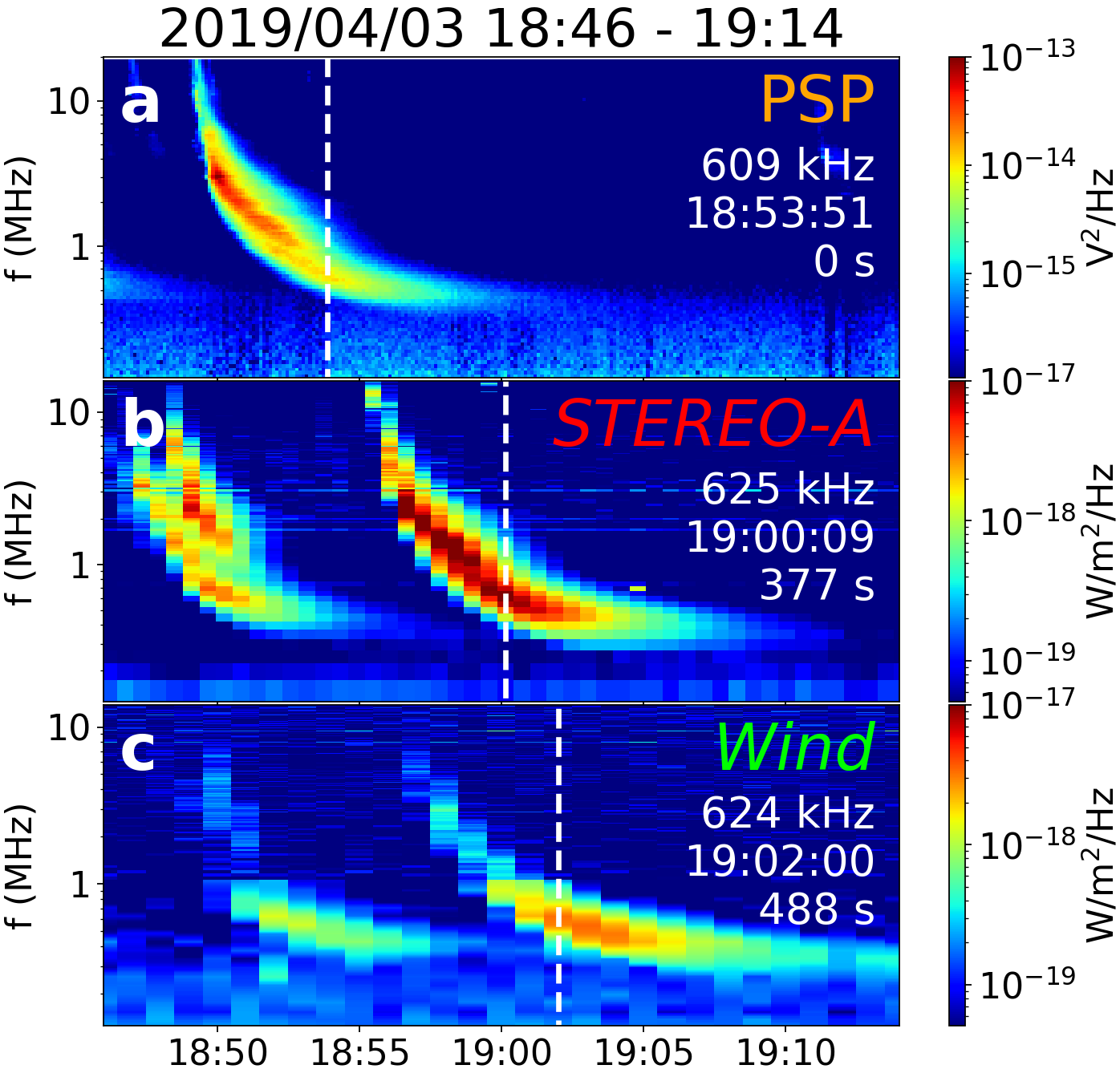}
\caption{Radio measurements of the 2019 April 3 type III burst. \replaced{The power spectral density  for RFS/HFR ($1.3$~MHz -- $19.2$~MHz) and RFS/LFR ($10.5$~kHz -- $1.7$~MHz).}{(a) The power spectral density $V_r^2$ for PSP/RFS. (b) The power flux density $S$ for \textit{STEREO-A}/WAVES. (c) The power flux density $S$ for \textit{Wind}/WAVES. White dashed lines indicate times of peak fluxes at 609~kHz, 625~kHz, and 624~kHz.}}
\label{fig:spectra}
\end{figure*}

\begin{figure}[htb]
\centering
\includegraphics[angle=0,width=0.9\linewidth]{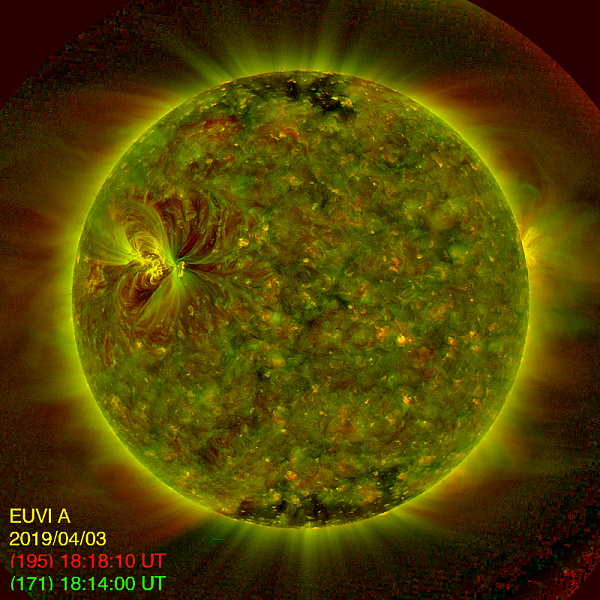}
\caption{Image of the Sun taken by the \textit{STEREO-A}/SECCHI/EUVI on 2019 April 3 at 18:14:00 ~UT ($171${\AA}) and 18:18:10~UT ($195${\AA}).
\replaced{The individual images are available at \href{http://sd-www.jhuapl.edu/secchi/wavelets}{http://sd-www.jhuapl.edu/secchi/wavelets}}{Enhanced images for individual wavelength channels are available at \href{http://sd-www.jhuapl.edu/secchi/wavelets/fits/201904/03/}{http://sd-www.jhuapl.edu/secchi/wavelets/fits/201904/03/}}.
An animation of just the SECCHI/EUVI ($195${\AA}) sequence is available. The animation field of view is zoomed in to the area around the flare in the upper right quandrant of the full solar image shown in the figure. The video begins on 2019 April 3 at 18:00:30 and ends the same day at 20:05:30. The realtime duration of the video is 2 seconds. The movie of the solar flare is derotated using the reference time corresponding to the first image (18:00 UT).}
\label{fig:euvi}
\end{figure}

\begin{figure*}[htb]
\centering
\includegraphics[angle=0,width=0.8\linewidth]{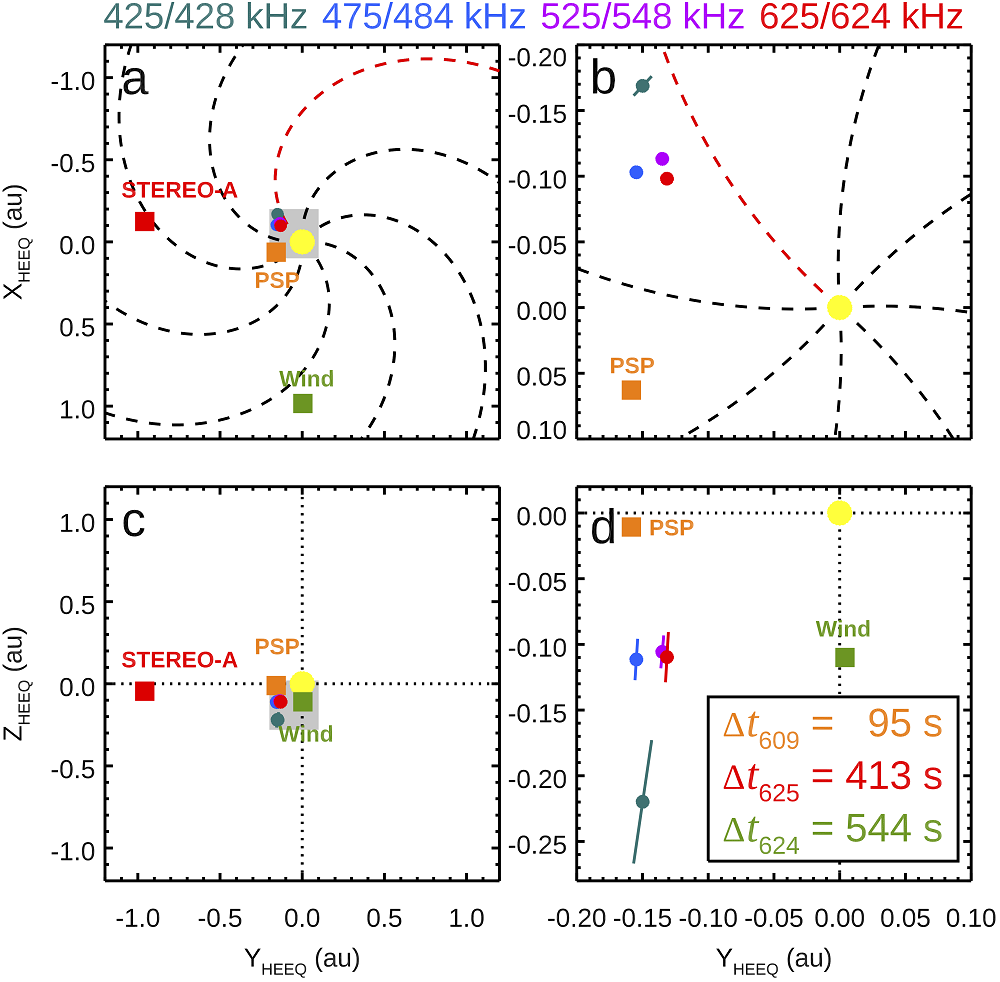}
\caption{Radio propagation analysis of the 2019 April 3 type III burst.
(a)--(d) Triangulated type III burst locations by \textit{STEREO-A} and \textit{Wind} in the XY$_{\rm{HEEQ}}$ (top) and ZY$_{\rm{HEEQ}}$ (bottom) planes. Colors denote frequencies shown on the top. Rectangles show spacecraft locations. Dashed lines indicate Parker spirals (a red one is rooted in the solar flare location). Grey areas in panels a and c show axis ranges in panels b and d, respectively.}
\label{fig:triang}
\end{figure*}

\begin{figure}[htb]
\centering
\includegraphics[angle=0,width=0.9\linewidth]{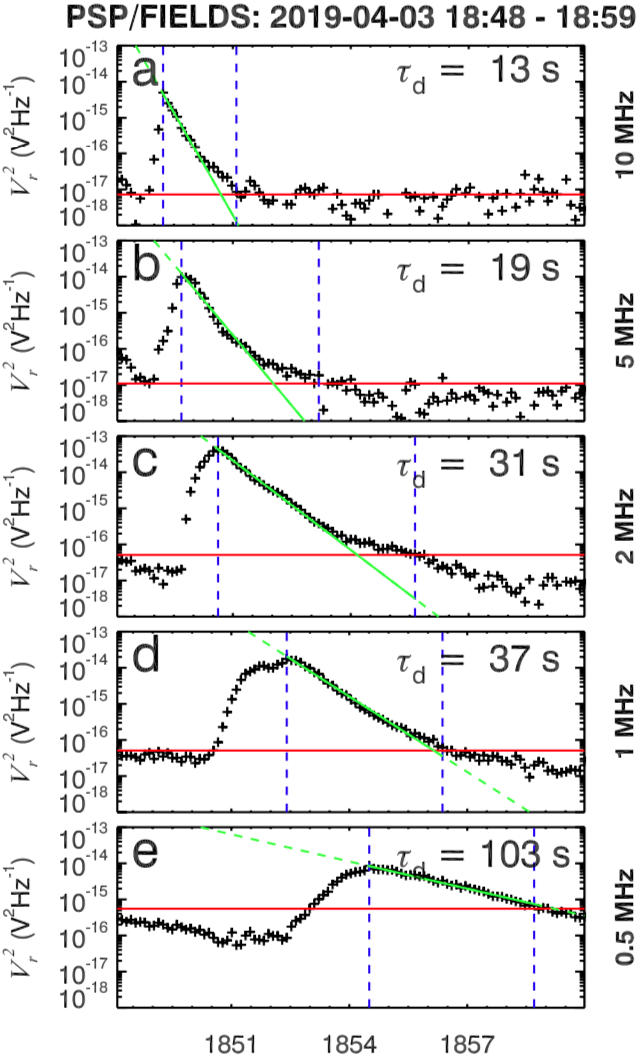}
\caption{Radio measurements of the 2019 April 3 type III burst.
(a)--(e) Fixed frequency light curves of the voltage power spectral density recorded by the RFS instrument for five frequency channels.
Red lines show median values in given time intervals.
Dashed blue lines denote peak fluxes and last points above median values.
Green lines show results of decay time fitting (equation~\ref{eq:fall}).}
\label{fig:lines}
\end{figure}

\begin{figure}[htb]
\centering
\includegraphics[angle=0,width=0.7\linewidth]{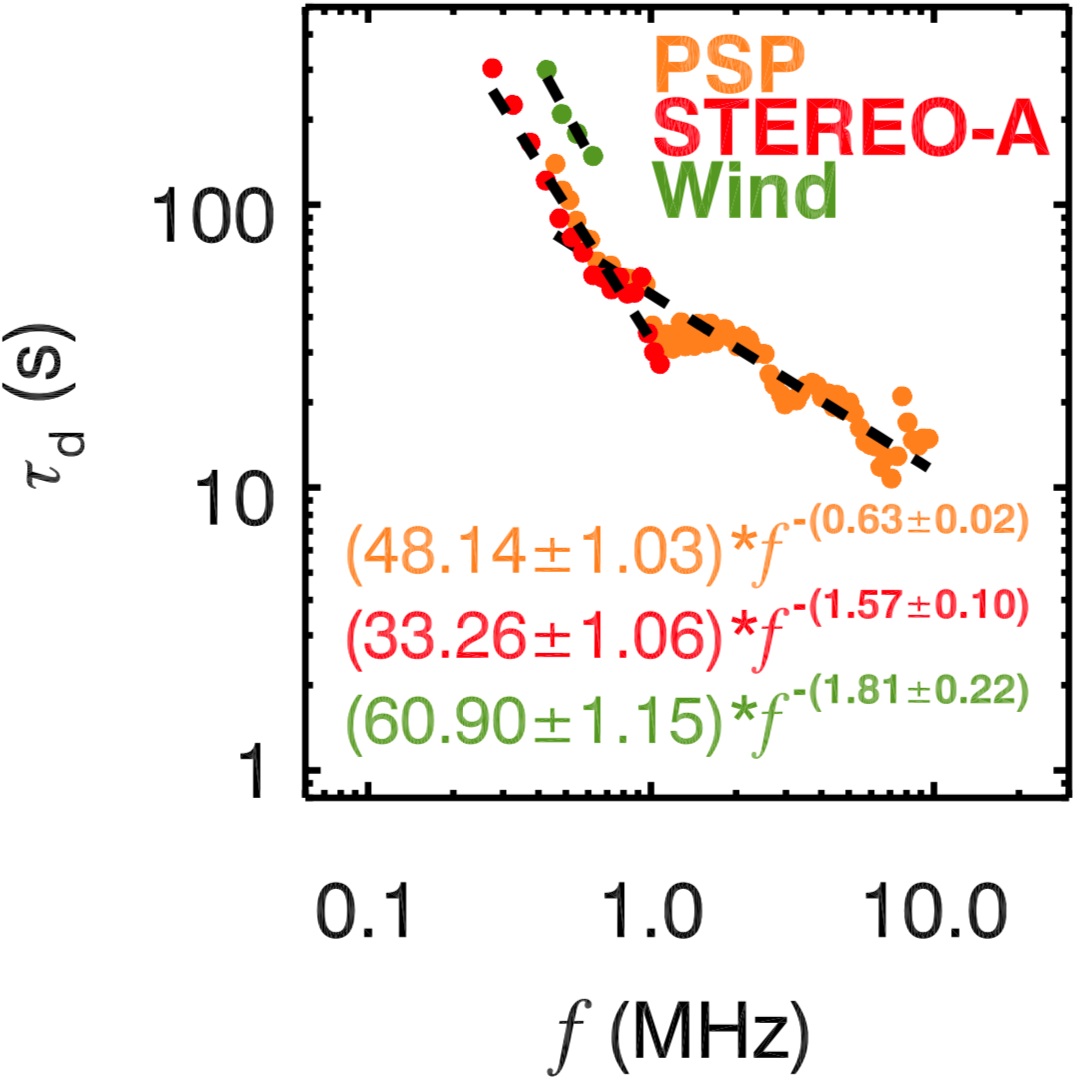}
\caption{Radio measurements of the 2019 April 3 type III.
Decay times $\tau_d$ for PSP, \textit{STEREO-A}, and \textit{Wind} as a function of frequency are shown in orange, red, and green, respectively.}
\label{fig:decay_case}
\end{figure}

\begin{figure}[htb]
\centering
\includegraphics[angle=0,width=0.9\linewidth]{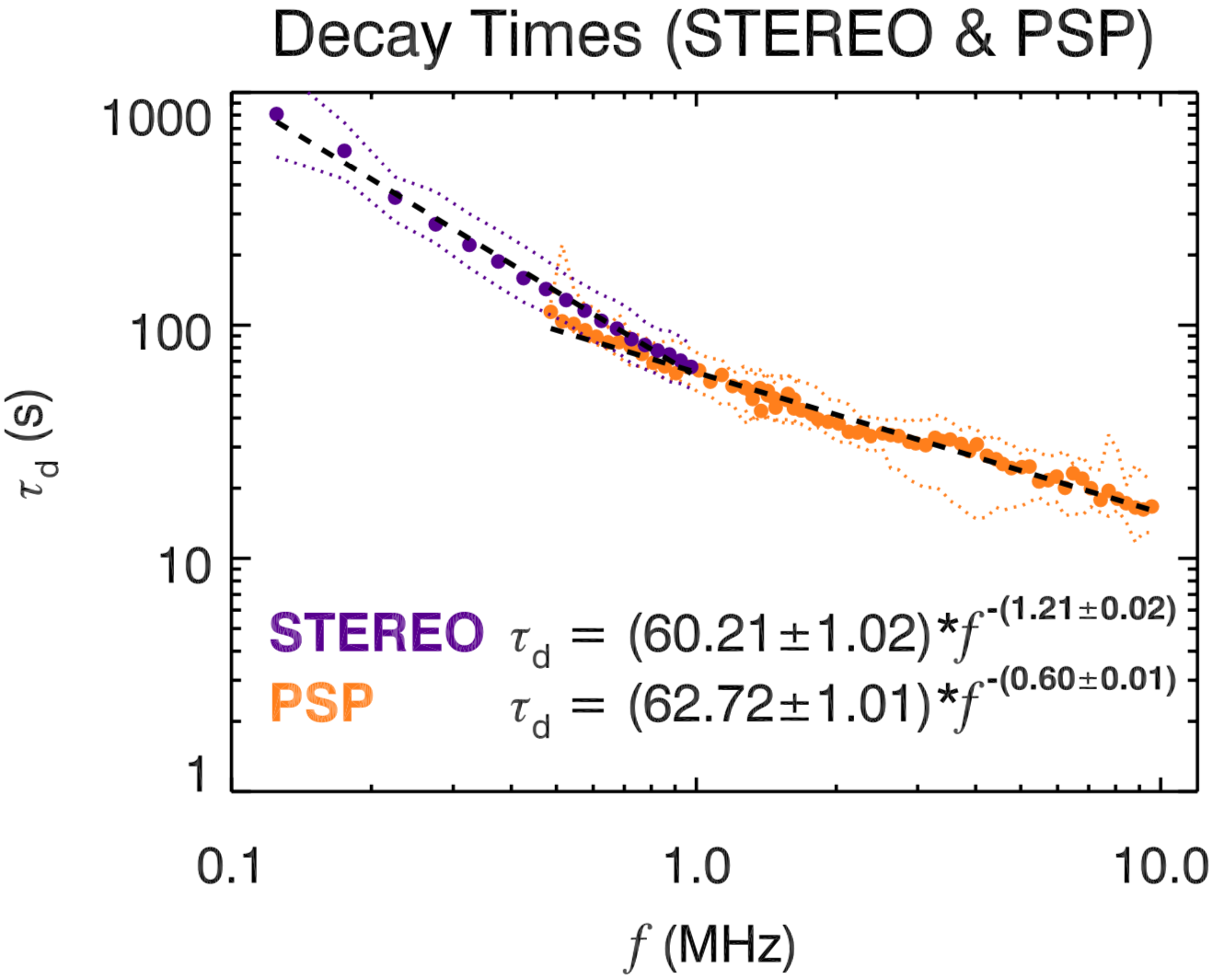}
 \caption{Results of the statistical survey of 152 and 30 type III radio bursts for \textit{STEREO} and PSP.
Median values of decay times $\tau_d$ for \textit{STEREO} and PSP as a function of frequency are shown in purple and orange, respectively.
Error bars are 25th/75th percentiles. Dashed black lines represent results of power-law fitting for the two data sets separately (equation~\ref{eq:power}).}
 \label{fig:fit_decay}
\end{figure}

\begin{figure}[htb]
\centering
\includegraphics[angle=0,width=0.9\linewidth]{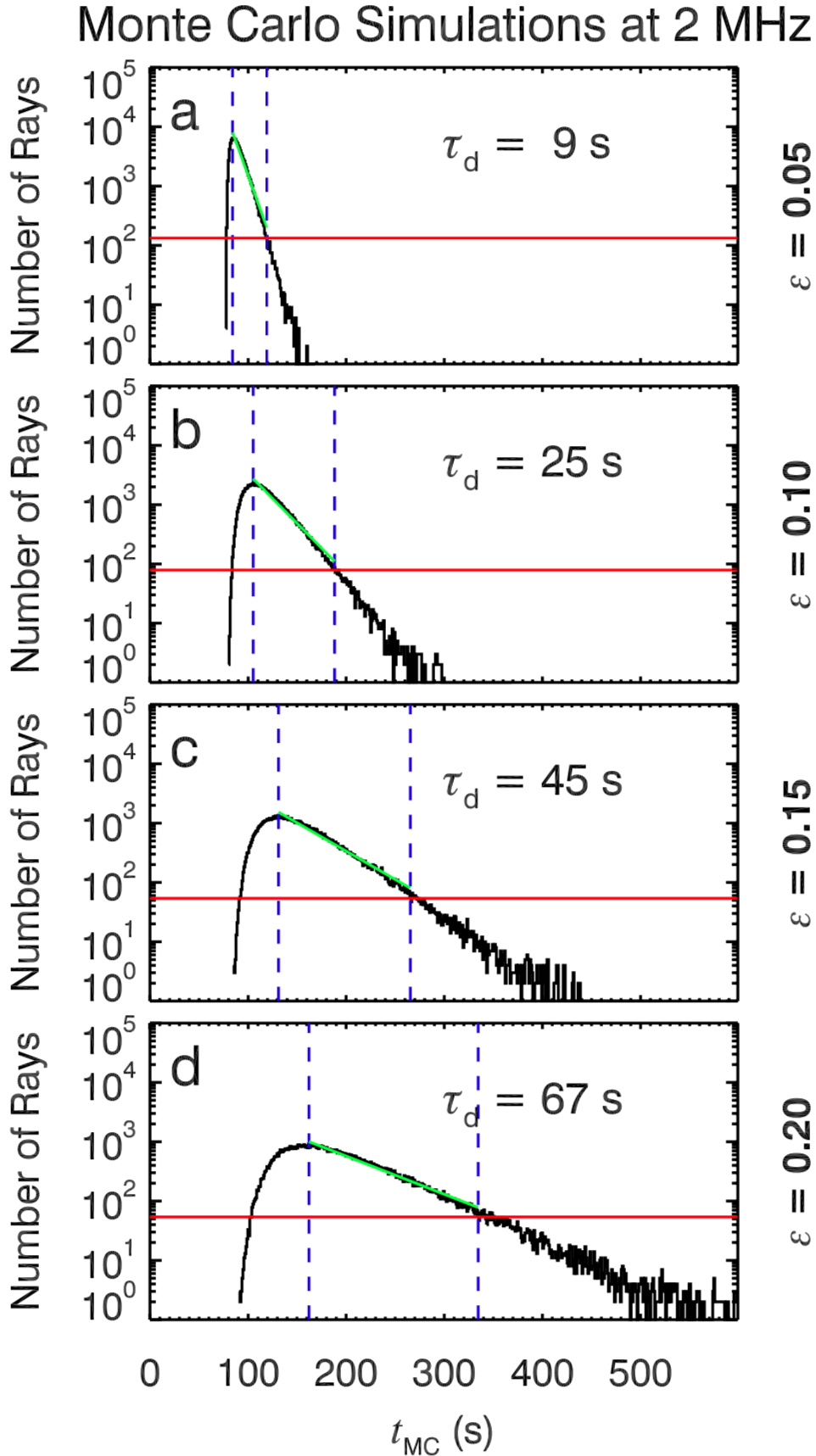}
\caption{Results of Monte Carlo simulations at $2$~MHz.
(a)--(d) Histograms of simulated time arrivals $t_{\rm{MC}}$
for various levels of relative electron density fluctuations $\epsilon$.
Red lines show median values.
Dashed lines denote peak fluxes and last points above median values.
Green lines show results of decay time fitting (Equation \ref{eq:fall}).}
 \label{fig:mc_decay}
\end{figure}

\begin{figure}[htb]
\centering
\includegraphics[angle=0,width=0.9\linewidth]{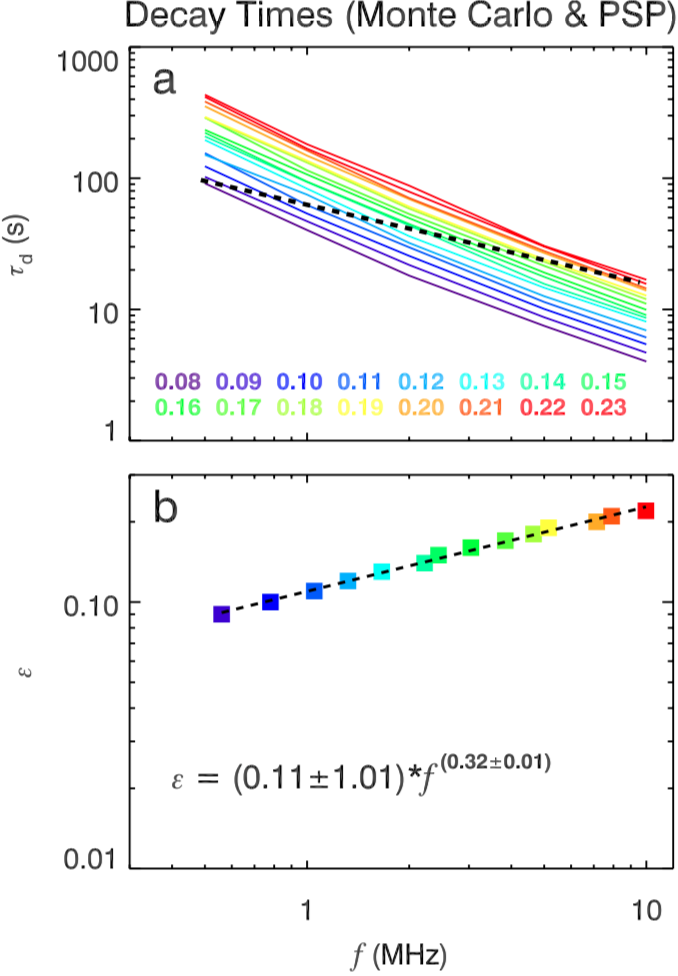}
 \caption{Comparison of Monte Carlo simulations and PSP observations.
 (a) Simulated decay times $\tau_d$ as a function of frequency for $16$~levels of relative density fluctuations $\epsilon$ in color.
 A dashed black line shows fitting results from PSP (Figure~\ref{fig:fit_decay}).
 (b) Relative density fluctuations $\epsilon$ as a function of frequency retrieved from intersections between Monte Carlo simulations and PSP observations. A dashed black line represents results of power-law fitting.}
 \label{fig:eps}
\end{figure}
\begin{figure}[htb]
\centering
\includegraphics[angle=0,width=0.99\linewidth]{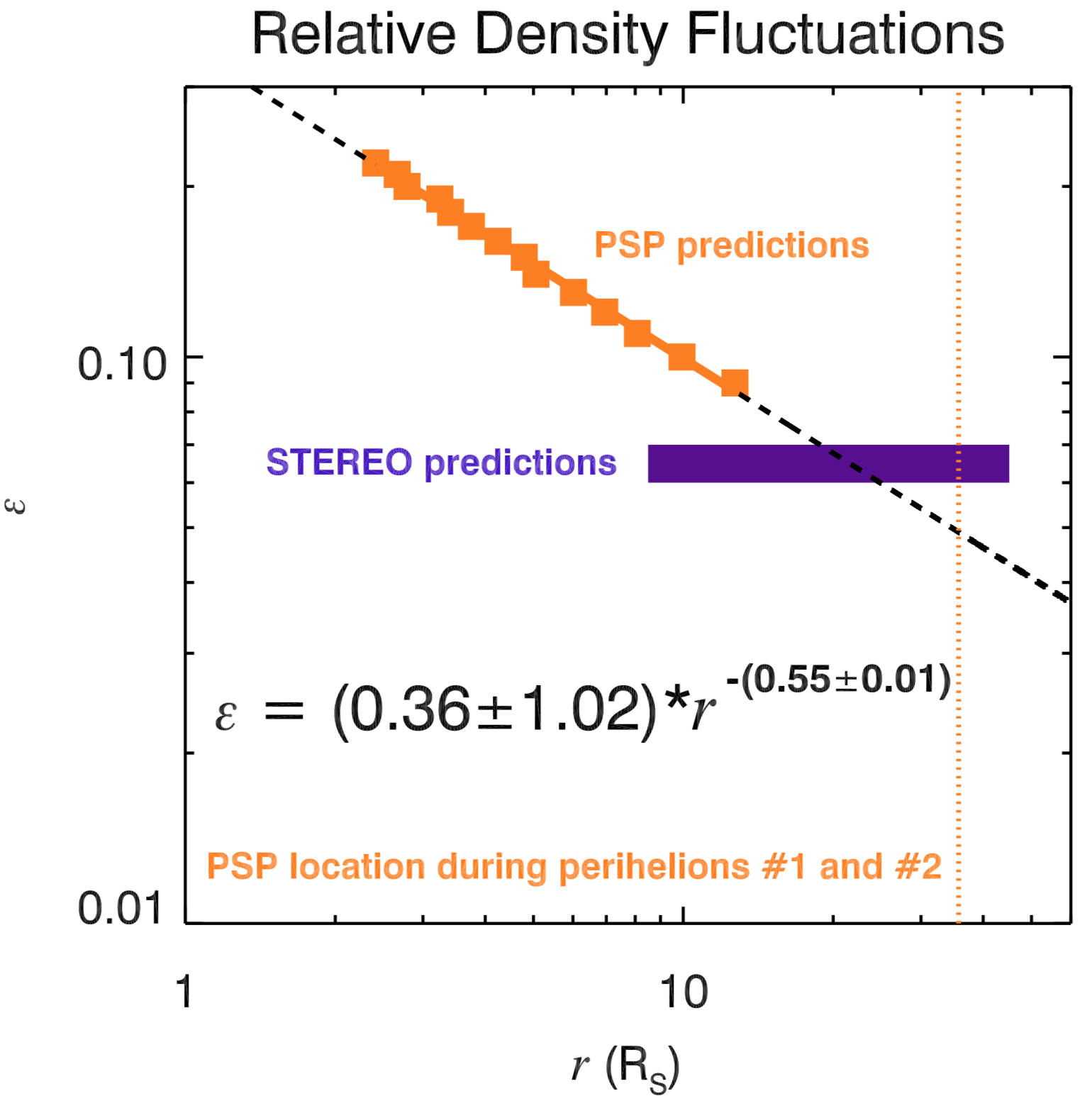}
 \caption{Results of Monte Carlo simulations and PSP observations.
Relative density fluctuations $\epsilon$ from Figure~\ref{fig:eps}b as a function of radial distance $r$ are denoted by orange squares. A solid black line represent results of power-law fitting. Predicted relative density fluctuations $\epsilon$ by \textit{STEREO} are shown in purple. A dotted orange line indicates radial distance of PSP during the Perihelion \#1 and \#2.}
 \label{fig:eps_r}
\end{figure}
\begin{figure*}[htb]
\centering
\includegraphics[angle=0,width=0.63\linewidth]{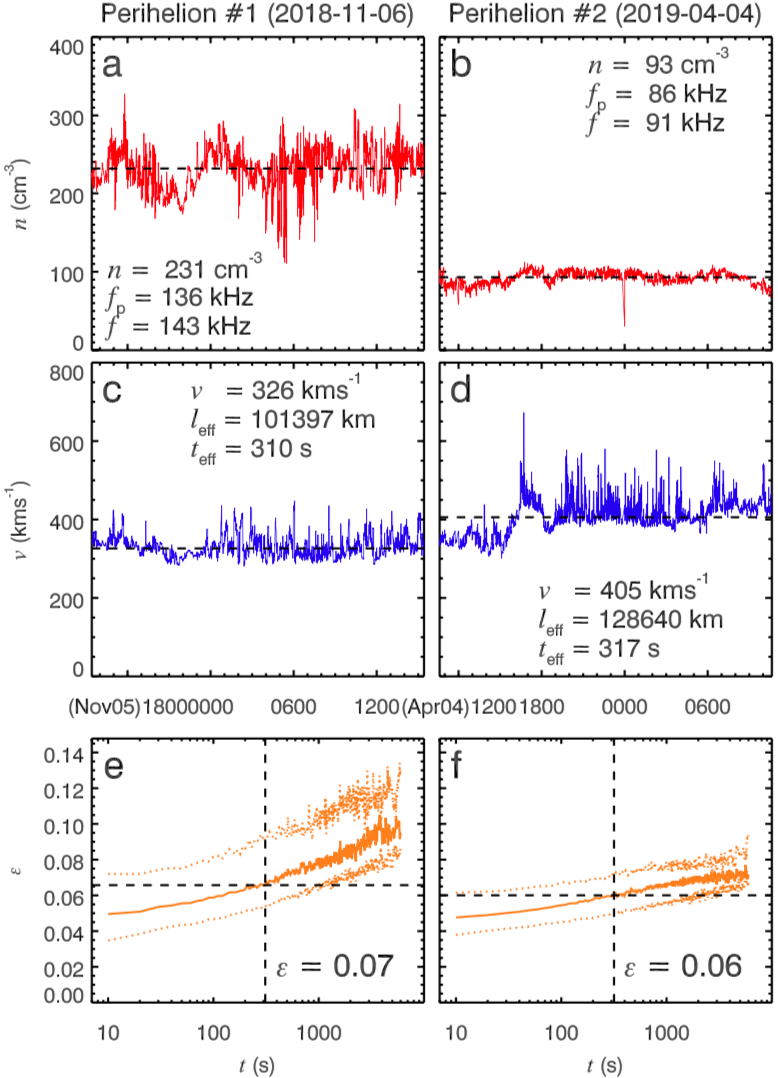}
 \caption{PSP plasma measurements $12$~hours before and $12$~hours after the perihelion~\#1 (left) and perihelion~\#2 (right).
(a,b) Plasma density.
(c,d) Bulk velocity. Dashed black lines show median values.
(e,f) Median values of relative density fluctuations $\epsilon$ as a function of temporal scale $t$. Error bars are 25th/75th percentiles.
Dashed black lines correspond to the effective scales of Monte Carlo simulations for frequency of $143$~kHz and $91$~kHz obtained from plasma parameters obtained during the perihelion~\#1 and perihelion~\#2.}
 \label{fig:sweap}
\end{figure*}

We performed a statistical analysis of $30$~type III radio bursts observed by PSP during the perihelion \#2 (2019~April~1 -- 2019~April~10).
During this period radial distances from the Sun ranged from $35.7$~$R_\Sun$ to $53.8$~$R_\Sun$. We included only intense, simple, and isolated emissions.
We show an analysis of a type III burst from 2019 April 3 when PSP was at $68^\circ$ east from a Sun-Earth line at $36.7$~$R_\Sun$ from the Sun as an example from our list of events.
Figure~\ref{fig:spectra}\added{a} displays the power spectral density $S$ from the RFS/HFR and RFS/LFR instruments using the average auto spectral data of the V1--V2 dipole input channel. PSP detected the type III  burst with an onset time at about 18:48 UT.
The type III burst was also measured by the \textit{Wind}/WAVES and \textit{STEREO-A}/WAVES instruments
\added{(Figures~\ref{fig:spectra}b and \ref{fig:spectra}c)}.
During this event, the \textit{Wind} spacecraft was on a Sun--Earth line at $0.99$~au from the Sun, whereas \textit{STEREO-A} was at $97^\circ$ east and $0.97$~au from the Sun.
\added{We have analyzed time delays between peak fluxes for close frequency channels of 609~kHz, 625~kHz, and 624~kHz for PSP, \textit{STEREO-A}, and \textit{Wind}, respectively. We selected these channels as higher frequencies were not observed by \textit{Wind}, while a PSP radio signal at lower frequencies was affected by QTN. The type III burst was delayed by $\delta t_{STA} = 377$~s, and $\delta t_{Wind} = 488$~s between PSP and \textit{STEREO-A} and \textit{Wind}, respectively.}

A solar flare triggering this emission has been located on the far side of the Sun from a view of the Earth. Hence we cannot retrieve its intensity and location as spacecraft embarking X-ray imagers orbit the Earth.
However, the active region has been observed by \textit{STEREO-A}/Sun Earth Connection Coronal and Heliospheric Investigation/Extreme Ultraviolet Imager \citep[SECCHI/EUVI; ][]{2008SSRv..136...67H}.
We have used the wavelet technique by \citet{2008ApJ...674.1201S} to produce a composite image for the  $171${\AA} and  $195${\AA} channels (Figure~\ref{fig:euvi}).
The coordinates for the footpoint of the loops where some activity is observed at $195${\AA} are $[-131^\circ, 6^\circ]$ in Stonyhurst-Heliographic longitude and latitude (see the Figure 2 animation).

The favorable configuration of \textit{Wind} and \textit{STEREO-A} allows us to accurately locate the sources of the type III bursts by radio triangulation \citep{2014SoPh..289.4633K,2016ApJ...823L...5K}.
We identified data points that correspond to peak fluxes for four pairs of frequency channels observed by \textit{Wind}/WAVES
($428$~kHz, $484$~kHz, $548$~kHz, and $624$~kHz) and \textit{STEREO-A}/WAVES
($425$~kHz, $475$~kHz, $525$~kHz, $625$~kHz) with signals above background levels.
We triangulated the radio sources using wave vector directions during these peak fluxes (Figure~\ref{fig:triang}).
\added{Specifically, we consider radio source location to be the closest point between the two wave vectors, and the shortest distance between the wave vectors indicates the error of triangulated source.}
We also included the Parker spiral rooted in the solar flare site assuming a \deleted{typical} solar wind speed of \replaced{500}{400} km~s$^{-1}$ to illustrate \replaced{the}{a possible} path followed by the electron beam \citep[a red dashed line; ][]{1958ApJ...128..677P}.
Generally, triangulated source regions of higher frequencies are closer to the Sun.
\added{Obtained error bars are noticeable only in the $\rm{XZ_{HEEQ}}$ plane due to the considerable smaller separation angle between \textit{STEREO-A} and \textit{Wind} in this plane (Figure~\ref{fig:triang}d).
We have also calculated light travel times between the triangulated radio source at 625/624~kHz and all three spacecraft: $\Delta t_{\rm{PSP}} = 95$~s, $\Delta t_{STA} = 413$~s, and $\Delta t_{Wind} = 544$~s. Using exclusively radio triangulation, we estimated the radio signal delays between PSP and \textit{STEREO-A} and \textit{Wind} to be
$\delta t_{STA}=\Delta t_{STA} - \Delta t_{PSP}=318$~s and $\delta t_{Wind}=\Delta t_{Wind} - \Delta t_{PSP}=449$~s, respectively. These values are comparable with actual delay signal measurements shown in Figure~\ref{fig:spectra} ($\Delta t_{STA} = 377$~s, and $\Delta t_{Wind} = 488$~s), which indicates that the radio triangulation technique provides reasonable source locations.}
Results of the triangulation confirm that the electron beam triggering the type III burst propagates roughly along the Parker spiral field near PSP.
The signal measured by \textit{STEREO-A}/WAVES was significantly stronger than that measured by \textit{Wind}/WAVES, which is consistent with the radio source located closer to \textit{STEREO-A}/WAVES.
Unfortunately, we are unable to perform the radio propagation analysis using PSP measurements as the PSP effective antenna parameters are not determined yet.

Figure \ref{fig:lines} shows fixed frequency light curves of the same event in four frequency channels ($0.5$~MHz, $1$~MHz, $2$~MHz, $5$~MHz, and $10$~MHz). The exponential decay of the power spectral density $S$ over several decades can be identified.
For the further analysis, we calculated median values of the power spectral density $S$
frequency by frequency to estimate background level (red lines in Figure \ref{fig:lines}).
We analyze data points between the peak time ($t_{\rm{peak}}$) and the last value above this level (\textit{i.e.,} between dashed blue lines in Figure \ref{fig:lines}). We assume an exponential decay profile of the power spectral density $S$, that can be described by following equation:

\begin{equation}
S(t)~=~\frac{I}{\tau_{\rm{d}}}\exp \Bigg(\frac{t_{\rm{peak}}-t}{\tau_{\rm{d}}}\Bigg),
  \label{eq:fall}
\end{equation}

where $t$ is the time, $t_{\rm{peak}}$ corresponds to the time of the peak power spectral density. Coefficients $I$ and
 $\tau_{\rm{d}}$ are parameters of
a gradient-expansion algorithm used to compute a non-linear least squares fit.
Figure \ref{fig:lines} shows results of this fitting for decay power spectral density profiles in green.

Figure \ref{fig:decay_case} shows type III burst decay times as a function of frequency for RFS, \textit{STEREO-A}/WAVES, and \textit{Wind}/WAVES.
We have achieved a very good agreement between RFS and \textit{STEREO-A}/WAVES for overlapping frequency channels (\textit{i.e.,} between $0.5$~MHz and $1$~MHz). However, decay times retrieved by \textit{Wind}/WAVES are considerably larger. It can be attributed to different emissions directivity due to relative spacecraft locations, when PSP and \textit{STEREO-A} are nearly along one Parker spiral, while \textit{Wind} is about $90^\circ$ away in the solar equatorial plane (Figure~\ref{fig:triang}a).
Next, we assume that the decay times $\tau_{\rm{d}}$ are frequency \replaced{depend}{dependent} as:
\begin{equation}  \label{eq:power}
\tau_{\rm{d}} (f)~=~\alpha f^{\beta}\mbox{.}
\end{equation}
This model \replaced{fit}{fits} the data well for all three spacecraft. We obtained the following spectral indices: $\beta_{\rm{PSP}}=(-0.63\pm0.02)$, $\beta_{STEREO-A}=(-1.57\pm0.10)$, and $\beta_{Wind}=(-1.81\pm0.22)$. These values were calculated by minimizing the $\chi^2$ error statistic with the 1-$\sigma$ uncertainty estimates. Despite variations in decay times between \textit{STEREO-A}/WAVES and \textit{Wind}/WAVES, the obtained spectral indices are rather similar. On the other hand, the $\beta_{\rm{PSP}}$ is significantly lager due to contributions by frequency channels between $1$~MHz and $10$~MHz, which are not covered by \textit{STEREO-A}/WAVES and \textit{Wind}/WAVES.

\begin{deluxetable}{llcc}
\tablecaption{The list of type III burst time-frequency intervals}
\tabletypesize{\scriptsize}
\tablehead{
\colhead{Date begin} & \colhead{Date end} &
\colhead{Frequency low} & \colhead{Frequency high} \\
\colhead{(UTC)} & \colhead{(UTC)} & \colhead{(MHz)} & \colhead{(MHz)}
}
\startdata
2019-04-01 01:57:00 & 2019-04-01 02:05:00 & 1.196 & 9.572 \\
2019-04-01 17:10:00 & 2019-04-01 17:17:00 & 0.545 & 8.428 \\
2019-04-01 20:25:00 & 2019-04-01 20:40:00 & 0.577 & 7.725 \\
2019-04-02 02:40:00 & 2019-04-02 02:57:00 & 0.809 & 9.572 \\
2019-04-02 04:44:00 & 2019-04-02 05:00:00 & 0.514 & 8.822 \\
2019-04-02 09:01:15 & 2019-04-02 09:12:00 & 0.764 & 9.572 \\
2019-04-03 04:00:00 & 2019-04-03 04:10:00 & 1.575 & 9.572 \\
2019-04-03 06:00:00 & 2019-04-03 06:10:00 & 1.566 & 9.572 \\
2019-04-03 09:20:00 & 2019-04-03 09:35:00 & 0.646 & 9.572 \\
2019-04-03 12:10:00 & 2019-04-03 12:25:00 & 0.514 & 9.572 \\
2019-04-03 12:35:00 & 2019-04-03 12:50:00 & 0.514 & 5.222 \\
2019-04-03 16:48:00 & 2019-04-03 17:00:00 & 0.514 & 9.572 \\
2019-04-03 17:00:00 & 2019-04-03 17:07:00 & 1.622 & 5.972 \\
2019-04-03 18:48:00 & 2019-04-03 19:00:00 & 0.514 & 9.572 \\
2019-04-03 21:05:00 & 2019-04-03 21:15:00 & 1.622 & 9.572 \\
2019-04-03 22:20:00 & 2019-04-03 22:40:00 & 0.514 & 9.572 \\
2019-04-04 02:35:00 & 2019-04-04 02:50:00 & 0.646 & 9.572 \\
2019-04-04 05:33:00 & 2019-04-04 05:45:00 & 0.764 & 9.572 \\
2019-04-04 22:10:00 & 2019-04-04 22:30:00 & 0.514 & 9.572 \\
2019-04-04 22:30:00 & 2019-04-04 22:40:00 & 0.646 & 9.572 \\
2019-04-05 03:25:00 & 2019-04-05 03:40:00 & 0.514 & 9.572 \\
2019-04-05 04:32:00 & 2019-04-05 04:42:00 & 0.855 & 9.572 \\
2019-04-05 10:52:00 & 2019-04-05 11:00:00 & 1.481 & 9.572 \\
2019-04-05 16:52:00 & 2019-04-05 17:15:00 & 0.514 & 9.572 \\
2019-04-05 17:06:00 & 2019-04-05 17:30:00 & 0.514 & 8.072 \\
2019-04-06 07:45:00 & 2019-04-06 07:59:00 & 0.514 & 9.572 \\
2019-04-06 09:40:00 & 2019-04-06 09:59:00 & 0.514 & 9.572 \\
2019-04-06 10:40:00 & 2019-04-06 10:59:00 & 0.514 & 9.572 \\
2019-04-07 09:50:00 & 2019-04-07 10:00:00 & 0.957 & 9.572 \\
2019-04-10 14:25:00 & 2019-04-10 14:45:00 & 1.566 & 9.572 \\
\enddata
\end{deluxetable}
\label{tab:events}

We performed the above-described analysis of the exponential decay times $\tau_{\rm{d}}$ on $30$~type III bursts observed by PSP during the perihelion $\#2$ case by case.
\added{We provide the list of type III burst time-frequency intervals, that can be used for further investigation by the community (Table~\ref{tab:events}).}
Figure~\ref{fig:fit_decay} displays median values of decay times $\tau_{\rm{d}}$ as a function of frequency.
We assume that the decay times $\tau_{\rm{d}}$ are frequency \replaced{depend}{dependent} as a power law (equation~\ref{eq:power}). The model \replaced{fit}{fits} the data very well. We obtained \replaced{spectral the}{the spectral} indices $\beta_{\rm{PSP}}$ of $-0.60\pm0.1$.
\deleted{Recently, \citet{2018ApJ...857...82K} performed a similar analysis of 152 type III bursts between $125$~kHz and $1$~MHz observed by the \textit{STEREO} spacecraft located at $1$~au.
The obtained spectral index is about twice smaller than for PSP ($\beta_{\rm{STEREO}}=-1.21\pm0.01$). We note that plasma frequency of $1$~MHz -- where the slope changes between \textit{STEREO} and PSP -- corresponds to a radial distance of $\sim8$~$R_\Sun$, where the solar wind speed typically exceeds the Alfv\'en speed, and the solar wind become superalfv\'enic. It is thus no surprise that type III burst properties change around frequency of $1$~MHz as the background plasma changes significantly.
We note that type III bursts exhibit also a maximum of power spectral density at $1$~MHz \citep{2014SoPh..289.3121K}.}

\subsection{Monte Carlo Simulations}
\label{sec:sim}

\citet{2007ApJ...671..894T} developed a Monte Carlo simulation code to investigate a role of refraction and scattering on propagation of \added{interplanetary} radio emissions \replaced{generated at $f~=~120$~kHz}{with isotropic sources}, when observed by spacecraft at $1$~au.
\deleted{\citet{2007ApJ...671..894T} used a set of six first-order differential equations derived by \citet{1963JATP...25..397H} to retrieve the position and direction vectors $\mathbf{R}$ and $\mathbf{T}$, respectively.
They launched $1,000$ rays from the isotropic point source located at altitudes with plasma levels corresponding to the $115$~kHz (the fundamental component; $f=1.05f_{\rm{pe}}$) and $60$~kHz (the harmonic component; $f=2f_{\rm{pe}}$).}
For the refraction, the solar wind electron density model of \citet{1984SoPh...90..401B} was used ($n\sim r^{-2.10}$).
For the scattering, \citet{2007ApJ...671..894T} assumed the power spectrum of electron density fluctuations in the solar wind $P_n$ in the inertial range to be proportional to the Kolmogorov spectrum.
\deleted{\citet{2007ApJ...671..894T} employed an empirical formula for the outer scale of the electron density fluctuations $l_o$ by \citet{2001SSRv...97....9W}, while the inner scale $l_i$ was assumed to be $100$~km \citep{1987sowi.conf...55M,1989ApJ...337.1023C}.}
The relative electron density fluctuations $\epsilon$ was set to be $0.07$.
\deleted{The scattering is introduced by adding a perturbation vector $\langle\mathbf{q}\rangle$ to $\mathbf{T}$ after each step when a ray suffers a regular refraction in a layer of thickness $\Delta S$ corresponding to the ray path length.\citet{2007ApJ...671..894T} assumed $\Delta S=10 l_{\rm{eff}}$ in the simulation code, where $l_{\rm{eff}}$ is the effective scale of electron density fluctuations defined as $l_{\rm{eff}}=l_i^{1/3} l_o^{2/3}$.
The vector $\langle\mathbf{q}\rangle$ is calculated from a Gaussian distribution of random numbers with a zero mean and a standard deviation of
These rays are traced until they reach $1$~au.}
We have modified the Monte Carlo technique of \citet{2007ApJ...671..894T} to simulate arrival times $t_{\rm{MC}}$ of radio emissions to $35.7$~$R_\sun$ (\textit{i.e.,} a radial distance of PSP during the perihelion \#2). Contrary to \citet{2007ApJ...671..894T}, we have used \deleted{a larger number of rays (100,000 rays instead of 1,000 rays),} a \added{ten times} finer simulation grid\deleted{ ($\Delta S= l_{\rm{eff}}$ \textit{vs.} $\Delta S=10 l_{\rm{eff}}$)},
\deleted{a} variable values of the inner scale $l_i$ \added{\citep{1989ApJ...337.1023C}}, and the \citet{1999ApJ...523..812S} density model, which works better for frequencies above $1$~MHz\deleted{We assumed a linear increase of $l_i$ with distance $R$ as $l_i=(R/R_{\Sun})$~km \citep{1987sowi.conf...55M,1989ApJ...337.1023C}}.

Figure \ref{fig:mc_decay} shows histograms of simulated arrival times $t_{\rm{MC}}$ of rays generated at $2$~MHz for four levels of the relative electron density fluctuations ($\epsilon = 0.05, 0.10, 0.15, 0.20$).
\added{We assumed a presence of the fundamental component only in accordance with \citet{2018ApJ...857...82K}.}
We identify similar exponential decay profiles as for the RFS measurements in Figure~\ref{fig:lines}.
We assume that the number of rays can be directly compared to the power spectral density $S$.
We have applied the same approach as for RFS data to derive the decay times $\tau_{\rm{d}}$ from these histograms (Figure \ref{fig:lines}).
We have found that the exponential model described by equation \ref{eq:fall} is in a good agreement with simulated data.
A direct comparison between PSP observations (Figure \ref{fig:lines}c) and Monte Carlo simulations (Figures \ref{fig:mc_decay}b and \ref{fig:mc_decay}c) suggests the relative electron density fluctuations $\epsilon$ at the effective turbulence scale length to be between $0.10$ and $0.15$ for emission generated at $2$~MHz.

Next, we performed the  Monte Carlo simulations for 5 frequency channels ($0.5$~MHz, $1$~MHz, $2$~MHz, $5$~MHz, and $10$~MHz), and $16$~levels of the relative electron density fluctuations $\epsilon$ between $0.08$ and $0.23$. Figure \ref{fig:eps}a shows simulated and observed decay times $\tau_{\rm{d}}$ \textit{vs.} frequency $f$.
While previous analysis of the \textit{STEREO} data by \citet{2018ApJ...857...82K} suggested that observed decay times $\tau_{\rm{d}}$ can be explained by scattering due to nearly constant relative electron density fluctuations ($\epsilon=0.06-0.07$), we need variable values to interpret the PSP data ($\epsilon=0.09-0.22$). Figure \ref{fig:eps}a displays relative density fluctuations $\epsilon$ as a function of frequency. The higher frequencies require larger levels of $\epsilon$ to explain observed decay times $\tau_{\rm{d}}$ by PSP. We have found that this relation can be described by a power law as:

\begin{equation}  \label{eq:power2}
\epsilon (f)~=~\alpha f^{\beta}\mbox{.}
\end{equation}

Next, we have converted frequencies to radial distances using the \citet{1999ApJ...523..812S} density model (Figure~\ref{fig:eps_r}). The obtained relation can be described by power-law type with a spectral index of $-0.55\pm0.01$ for radial distances from $2.4$~$R_\sun$ up to $13.9$~$R_\sun$. We note that \textit{STEREO} results between $125$~kHz and $1$~MHz (\textit{i.e.,} from $8.4$~$R_\sun$ up to $45.1$~$R_\sun$) suggest nearly constant relative density fluctuations $\epsilon = 0.06-0.07$ at the effective turbulence scale length \citep{2018ApJ...857...82K}.

\subsection{Density Fluctuations}
\label{sec:fluc}

Finally, we have compared predicted relative density fluctuations by \textit{STEREO} \citep{2018ApJ...857...82K}, and the measured ones by PSP/SWEAP during the perihelions \#1 and \#2. Specifically, we use density and velocity measurements based on the proton moments from the SPC Faraday cup.
We have calculated median values of plasma density and bulk velocity during periods $12$~hours before and $12$~hours after the closest approaches (Figures~\ref{fig:sweap}a--\ref{fig:sweap}d). Obtained plasma densities corresponds to local plasma frequencies $f_{\rm{p}}=137$~kHz and $f_{\rm{p}}=86$~kHz for the perihelions \#1 and \#2, respectively. The Monte Carlo technique assumes the fundamental  emission to be generated at $1.05f_{\rm{p}}$ resulting in
$f=143$~kHz and $f=91$~kHz. Next, we retrieve effective spatial scales $l_{\rm{eff}}$ for these frequencies using empirically derived model of the inner and outer scales of the electron density fluctuations. We compared these spatial scales with median values of plasma bulk velocities to obtain effective temporal scales of the density turbulence $t_{\rm{eff}}$.
Next, we have calculated relative density fluctuations $\epsilon$ as a function of the time scale $t$ between $10$~seconds and $100$~minutes:

\begin{equation}
\epsilon(t)~=~\frac{\langle |n - \langle n \rangle_t |\rangle_t}{\langle n\rangle_t}
  \label{eq:eps2}
\end{equation}

For time scales corresponding \added{to} the effective turbulence scale length in our Monte Carlo simulation technique we obtained $\epsilon=0.07$ and $\epsilon=0.06$ for the perihelions \#1 and \#2, respectively.
It is in a very good agreement with relative density fluctuations $\epsilon = 0.06-0.07$ predicted by \citet{2018ApJ...857...82K}.

\section{Discussion and Summary}
\label{sec:con}

\replaced{PSP}{While type III bursts have been been observed for almost $70$~years, a proper model to explain their obscure properties is still missing. PSP/RFS is the state-of-the-art instrument which} allows us to investigate \replaced{type III bursts generated in the solar wind near the Sun}{interplanetary solar radio bursts} with an unprecedented time \deleted{and frequency}resolution
\added{near the Sun}.
\added{For the first time, we can accurately retrieve type III burst decay times for frequencies between $1$~MHz and $10$~MHz to remotely probe solar wind turbulence near the Sun.
Although PSP/RFS accumulated wealth of data during the first two perihelions, type III bursts were almost exclusively observed during the perihelion \#2 only despite the ongoing solar minimum.}

We show an analysis of a type III burst that occurred on 2019 April 3 during the perihelion \#2 (\replaced{Figures \ref{fig:spectra} and \ref{fig:euvi})}{Figure~\ref{fig:spectra}), which was associated with the active region at $[-131^\circ, 6^\circ]$ in Stonyhurst-Heliographic longitude and latitude (Figure~\ref{fig:euvi})}.
\added{We note that this active region was responsible for a majority of solar activity during the perihelion \#2.
The simple type III burst was also observed couple minutes later by \textit{STEREO-A} and \textit{Wind}, which allowed us to compare signal delays between the three spacecraft.}
We have localized radio sources using triangulation technique applied to \textit{STEREO-A}/Waves and \textit{Wind}/Waves measurements (Figure \ref{fig:triang}).
\added{Triangulated radio sources lie near the modeled Parker spiral rooted in the active region. The results from the radio triangulation and time delay analysis confirm that this radio emission is related to the active region.}

\replaced{We have found that the power spectral density $S$ decreases exponentially over several decades at all three spacecraft (Figures \ref{fig:lines} and \ref{fig:decay_case}).}{We have analyzed RFS fixed frequency light curves for five frequency channels (Figure \ref{fig:lines}; $0.5$~MHz, $1$~MHz, $2$~MHz, $5$~MHz, and $10$~MHz). We observed the characteristic exponential decay profile for all frequency channels.
However, the fit does not perform well for late phases at higher frequencies and a clear hardening of the profile can be recognized.
This effect is probably related to underestimating of background level for the exponential decay fit at higher frequencies. While the type III burst is above the estimated background level for almost six minutes at $0.5$~MHz, it is only around two minutes at $10$~MHz.
Consequently, the background level at $10$~MHz on the same 11 minute time interval is relatively lower when compared to measurements at $0.5$~MHz.
Other explanation would be a presence of the harmonic component and/or another weaker type III burst.
Nevertheless, these deviations in late phases are of minor importance to affect calculated decay times since obtained values predominantly rely on data points succeeding peak fluxes, where the exponential decay fit performs very well.}

\added{Figure~\ref{fig:decay_case} shows a comparison of decay times observed by all three spacecraft. Despite different radial distances of PSP and \textit{STEREO-A} from the Sun, obtained results are comparable for overlapping frequencies as the two spacecraft lie approximately on the same Parker spiral. It indicates that scattering -- if responsible for long exponential decays -- occurs primarily near sources, and radio waves propagate along straight lines afterwards. Nonetheless, the exponential decay exhibits a clear hardening above $1$~MHz, which will be discussed later. We note that the hardening would be even more pronounced if the late phases in Figure~\ref{fig:lines} are included.}

\added{On the other hand, decay times observed by \textit{Wind} are considerably longer perhaps due to larger longitudinal separation with the active region. However, the slope of the power-law fit is similar to \textit{STEREO-A}.
A comparison of decay times from widely separated spacecraft may provide additional information to radio triangulation and/or time delay analysis to complement radio source localization.}

We have investigated a large number of type III bursts in order to statistically retrieve their exponential decay times $\tau_{\rm{d}}$ as a function of frequency $f$ (Figure \ref{fig:fit_decay}). Using the power-law model we obtain a spectral index $\beta_{\rm{PSP}}$ of $-0.60\pm0.01$.
\added{Recently, \citet{2018ApJ...857...82K} performed a similar analysis of 152 type III bursts between $125$~kHz and $1$~MHz observed by the \textit{STEREO} spacecraft located at $1$~au.
The obtained spectral index is about twice smaller than for PSP ($\beta_{\rm{STEREO}}=-1.21\pm0.01$).
However, statistical results between $0.5$~MHz and $1$~MHz by both PSP (30 events detected in April 2019 at $\sim0.17$~au) and \textit{STEREO} (152 events measured between May 2007 and February 2013 at $\sim1$~au) are comparable. If one assumes that exponential decay is caused by scattering, then it confirms that scattering is important only close to radio sources and later type III bursts propagate along straight lines.}

\added{The \citet{1999ApJ...523..812S} density model suggests that $1$~MHz -- where the slope changes between \textit{STEREO} and PSP -- corresponds to a radial distance of $\sim8$~$R_\Sun$ (the fundamental component) or $\sim14$~$R_\Sun$ (the harmonic component), where the solar wind speed typically exceeds the Alfv\'en speed, and the solar wind become superalfv\'enic -- the solar wind is no longer in contact with the Sun since Alfv\'en waves cannot travel back to the Sun
Type III burst properties change around $1$~MHz as the ambient plasma evolves significantly. Moreover, type III bursts statistically exhibit also a maximum of power spectral density around $1$~MHz \citep{2014SoPh..289.3121K}.
Furthermore, it is also possible that we rather observe the fundamental component below $1$~MHz, while the harmonic component is dominant above $1$~MHz. If it is the case, variations in exponential decay times within one single type III burst can be used to distinguish the component one from another.}

We have implemented a Monte Carlo simulation technique to study a role of scattering to \added{type III burst} decay times (Figure \ref{fig:mc_decay}).
\added{We assumed a presence of the fundamental component only since the used Monte Carlo simulation technique does not perform well for the harmonic one. Specifically, \citet{2018ApJ...857...82K} showed that distributions of simulated arrival times of the harmonic component are very narrow with short onset times, which is inconsistent with type III burst observations.
Moreover, following assumptions have been included in the Monte Carlo code: (1) an isotropic point source, (2) the \citet{1999ApJ...523..812S} density model, (3) a power-law distribution of density fluctuations, (4) empirically modeled inner and outer scales, (5) a constant value of isotropic density fluctuations. Obviously, these simplifications affect our analysis and can be improved in the future. For example, a finite size dipole/quadrupole emission pattern would probably slightly increase modeled decay times. A variable anisotropic density fluctuation model would work better for radio emission generated near the Sun.}

From the arrival times we calculated the decay times $\tau_{\rm{d}}$ that we compare to those observed by RFS (Figure~\ref{fig:eps}).
\added{As scattering plays a significant role near radio sources only, we may assume variable levels of $\epsilon$ when comparing PSP observations with Monte Carlo simulations for each frequency separately.}
Our results suggest that the exponential decay of the observed power spectral density can be explained by the scattering of radio signal by density inhomogeneities in the solar wind.
Obtained relative electron density fluctuations $\epsilon$ are $0.09-0.22$ at the effective turbulence scale length. We note that this range depends on our assumptions of the effective scale $l_{\rm{eff}}$ of the electron density fluctuations near actual radio sources.

\added{Predicted electron density fluctuations increase closer to the Sun below the Alfv\'en point (Figure~\ref{fig:eps_r}).
Nonetheless, \textit{STEREO} observations indicate constant density fluctuations above the Alfv\'en point \citep{2018ApJ...857...82K}.
Possible explanation of this discrepancy would be that solar wind turbulence is primarily formed near the Sun, while it remains frozen into the solar wind once beyond the Alfv\'en point.}

Finally, we have analyzed plasma parameters measured \textit{in situ} by PSP during the perihelion \#1 and the perihelion \#2 \added{to exploit unique observations near the Sun} (Figure \ref{fig:sweap}). Our results suggest that relative density fluctuations $\epsilon$ are $0.06-0.07$ at the effective turbulence scale length in our Monte Carlo simulation technique, which confirm previous predictions by \citet{2018ApJ...857...82K}.

\added{The main results of this study have been obtained by a statistical analysis of 30 type III bursts observed by PSP during the perihelion \#2, by Monte Carlo modeling of radio wave propagation in the solar wind, and by an analysis of \textit{in situ} plasma measurements during the perihelion \#1, and the perihelion \#2.
We have concluded that:}

\added{
\begin{enumerate}
	\item Type III burst decay times between $0.5$~MHz and $1$~MHz are statistically comparable at $\sim0.17$~au and $\sim1$~au, which confirms that scattering plays a major role in radio wave propagation near sources only.
	\item Type III burst decay times between $1$~MHz and $10$~MHz are statistically longer than expected based on previous observations at lower frequencies. I can be explained either by different ambient plasma parameters above the Alfv\'en point or that we observe preferably the harmonic component above $1$~MHz.
	\item If the latter is true, variations in exponential decay times can be used to distinguish fundamental and harmonic components within one single type III burst.
	\item By comparing PSP observations and Monte Carlo simulations, we predicted relative density fluctuations $\epsilon$ at radial distances between $2.5$~$R_\sun$ and $14$~$R_\sun$ to range from $0.22$ and $0.09$.
	\item Observed relative density fluctuations $\epsilon$ at a radial distance from the Sun of $35.7$~$R_\sun$ were $0.06-0.07$.
\end{enumerate}
}

\added{Note, however, that predicted relative density fluctuations $\epsilon$ are based on an assumption that we primarily observe the fundamental component of type III bursts only as the used Monte Carlo technique does not perform well for the harmonic component \citep{2018ApJ...857...82K}.}

\acknowledgements
The authors would like to thank the many individuals and institutions who contributed to making PSP, \textit{STEREO}, and \textit{Wind} possible. V.K. acknowledges support by an appointment to the NASA postdoctoral program at the NASA Goddard Space Flight Center administered by Universities Space Research Association under contract with NASA and the Czech Science Foundation grant 17-06818Y. O.K. thanks support of the Czech Science Foundation grant 17-06065S. E.P.K. was supported by an STFC consolidated grant ST/P000533/1. S.D.B. acknowledges the support of the Leverhulme Trust Visiting Professorship program.
\added{Data access and processing was performed using SPEDAS \citep{2019SSRv..215....9A}. All data is publicly available at \href{https://spdf.gsfc.nasa.gov/}{https://spdf.gsfc.nasa.gov/} and
\href{https://umbra.nascom.nasa.gov/}{https://umbra.nascom.nasa.gov/}.}


\clearpage


\end{document}